\documentclass[%
superscriptaddress,
a4paper,
amsmath,amssymb,
aps,
prd,
]{revtex4-2}

\usepackage{dcolumn}
\usepackage{bm}
\usepackage{graphicx}
\usepackage[colorlinks, linkcolor=blue, anchorcolor=blue, citecolor=blue]{hyperref}
\usepackage{subfigure}
\usepackage{multirow}
\usepackage{acronym}
\usepackage{tabularx}
\usepackage{booktabs}

\usepackage{color}

\begin{document}

\title{Premerger detection of massive black hole binaries using deep learning}

\author{Wen-Hong Ruan}
\email{ruanwenhong@ucas.ac.cn}
\affiliation{School of Fundamental Physics and Mathematical Sciences, Hangzhou Institute for Advanced Study, University of Chinese Academy of Sciences, Hangzhou 310024, China}
\affiliation{School of Physical Sciences, University of Chinese Academy of Sciences, No.19A Yuquan Road, Beijing 100049, China}

\author{Zong-Kuan Guo}
\email{guozk@itp.ac.cn}
\affiliation{CAS Key Laboratory of Theoretical Physics, Institute of Theoretical Physics, Chinese Academy of Sciences, Beijing 100190, China}
\affiliation{School of Physical Sciences, University of Chinese Academy of Sciences, No.19A Yuquan Road, Beijing 100049, China}
\affiliation{School of Fundamental Physics and Mathematical Sciences, Hangzhou Institute for Advanced Study, University of Chinese Academy of Sciences, Hangzhou 310024, China}

\begin{abstract}
Coalescing massive black hole binaries (MBHBs) are one of primary sources for space-based gravitational wave (GW) observations.
The mergers of these binaries are expected to give rise to detectable electromagnetic (EM) emissions with a narrow time window.
The premerger detection of GW signals is vital for follow-up EM observations.
The conventional approach for searching GW signals involves high computational costs.
In this study, we present a deep learning model to search for GW signals from MBHBs.
Our model is able to process 4.7 days of simulated data within 0.01 seconds and detect GW signals several hours to days before the final merger.
The model provides the possibility of the coincident GW and EM detection of MBHBs.
\end{abstract}

\maketitle

\section{Introduction}
\label{sec:Intro}
Since the first detection of gravitational wave (GW) signals from compact binary coalescence in 2015, the ground-based GW detectors have discovered an increasing number of GW events~\cite{LIGOScientific:2018mvr,LIGOScientific:2020ibl,LIGOScientific:2021usb,KAGRA:2021vkt}. 
These detectors are primarily sensitive to GW signals in the tens of hertz to kilohertz frequency range. 
Moreover, stochastic GW backgrounds in the nanohertz frequency range are detectable via pulsar timing arrays~\cite{NANOGrav:2020bcs,NANOGrav:2023gor,Reardon:2023gzh}.
Nonetheless, there remain extensive frequency ranges awaiting exploration.
By around 2030, several space-based GW detectors, including the Laser Interferometer Space Antenna (LISA)~\cite{amaro2017laser}, Taiji~\cite{hu2017taiji}, and TianQin~\cite{luo2016tianqin}, are scheduled to be launched, opening a window in the millihertz frequency range for GW observations. 

It is known that there exists massive black holes at the centers of galaxies~\cite{Kormendy:1995er,Magorrian:1997hw}.
Massive black hole binaries (MBHBs) are thought to form as galaxies evolve over time~\cite{Begelman:1980vb}. 
For space-based detectors, they are a highly significant class of GW sources~\cite{amaro2017laser,Ruan:2018tsw}.
We can test gravity in the highly nonlinear strong-field regime with GW signals from them~\cite{Ryan:1997hg,Collins:2004ex,Berti:2005ys,Cornish:2006ms}.
Furthermore, the coincident GW and electromagnetic (EM) detections for MBHBs will inaugurate a new era of multimessenger astrophysics, facilitating the comprehension of galaxy evolution at high redshifts~\cite{Cornish:2006ms}.
It also provides a new perspective for resolving Hubble tension through standard sirens~\cite{Holz:2005df,Chen:2017rfc,Wang:2021srv,Jin:2023sfc}.
There is anticipation of observing a variety of EM emissions associated with the environment surrounding MBHBs~\cite{Bogdanovic:2021aav}.
However, some of them have short observational timescales~\cite{Bogdanovic:2021aav,Paschalidis:2021ntt,Bode:2011tq,Palenzuela:2010nf,DalCanton:2019wsr}, leading to the demand for low-latency data analysis in GW observations.
Specifically, EM emissions from minidisks gradually disappear before the merger of MBHBs due to the shrinkage of minidisks, while they can provide a method for measuring spins of massive black holes~\cite{Paschalidis:2021ntt}.
Moreover, EM emissions from hot accretion flows~\cite{Bode:2011tq} or magnetized circumbinary disks~\cite{Palenzuela:2010nf,DalCanton:2019wsr} exhibit a transient peak around the merger of MBHBs.
Therefore, the premerger detection of GW signals facilitates prompt localization of the MBHB, enabling detailed planning for the observation of EM emissions. 

Currently, the commonly used approach for identifying GW signals amidst noise relies on matched filtering techniques~\cite{Owen:1998dk}, which have demonstrated remarkable efficacy in ground-based GW observations~\cite{LIGOScientific:2016vbw,LIGOScientific:2016sjg,LIGOScientific:2017bnn}. 
However, it is computationally expensive due to the requirements of calculating a detection statistic, such as the signal-to-noise ratio (SNR)~\cite{helstrom2013statistical}, for numerous waveform templates.
To search for GW signals from compact binary coalescence, a template bank containing about 250,000 waveform templates was employed in the first observational run of Advanced LIGO~\cite{Roy:2017oul}.
Moreover, it is estimated that around $10^{13}$ templates are required to search for GW signals from non-spinning MBHBs~\cite{Cornish:2006dt,Cornish:2006ms}.
A gridless search based on Metropolis-Hastings sampling and simulated annealing can be considered to significantly reduce the frequency of calculating the detection statistic~\cite{Cornish:2006dt}.
Nevertheless, in light of the demand to observe EM counterparts of MBHBs and further reduce the computational costs, there is considerable value in developing novel techniques for identifying GW signals.

Deep learning presents a great potential in GW data analysis due to its capacity for automatic feature extraction and complex pattern recognition.
Deep learning models have been extensively studied for the identification of GW signals~\cite{Gabbard:2017lja,George:2016hay,George:2017pmj,Gebhard:2019cov,Krastev:2019koe,
Wang:2019zaj,Dreissigacker:2019edy,Morawski:2019awi,Chan:2019fuz,Schafer:2020kor,Xia:2020vem,Cuoco:2020ogp,Dreissigacker:2020xfr,alvares2021exploring,Dodia:2021det,Schafer:2021fea,Antelis:2021qak,Yamamoto:2021use,
Ma:2022esx,Fan:2022wio,Szczepanczyk:2022urr,Choudhary:2022yje,Schafer:2022dxv,Nousi:2022dwh,Yamamoto:2022kuh,Meijer:2023yhn,Joshi:2023hpx,Zhao:2023ncy,Ma:2023ctz,Iess:2023quq,Yun:2023aqa,
Ruan:2021fxq,Zhao:2022qob}.
Specifically, the models designed in Ref.~\cite{Ruan:2021fxq,Zhao:2022qob} are capable of identifying GW signals from coalescing MBHBs.
However, these models are tailored for MBHB signals that have reached the ringdown phase and Ref.~\cite{Zhao:2022qob} does not account for the confusion noise generated by numerous galactic binaries~\cite{Nelemans:2001hp}.
In this work, we present a novel model based on residual network (ResNet)~\cite{He_2016_CVPR} and Transformer architecture~\cite{NIPS2017_3f5ee243}, named RTGW (ResNet-Transformer for Gravitational Waves), to identify MBHB signals prior to the final merger.
The architecture of ResNet has been widely used in neural networks designed for identification of GW signals~\cite{Dreissigacker:2019edy,Dreissigacker:2020xfr,alvares2021exploring,Dodia:2021det,Nousi:2022dwh,Yamamoto:2022kuh,Meijer:2023yhn,Joshi:2023hpx,Zhao:2023ncy}.
The galactic confusion noise is considered in the training and testing phases of our model.
The RTGW model has the ability to analyze data segments spanning several days, extracted from the detector strain data, within 0.01 seconds, enabling near real-time data analysis during detector operation.
It can be integrated as a search module into low-latency data analysis pipelines, such as the pipelines proposed in Ref.~\cite{Cornish:2021smq,Katz:2021uax,Chen:2023qga}, to efficiently identify MBHB signals before the final merger at low cost.
In this case, rapid localization of the MBHB can be achieved through the analysis of GW data, which guides the follow-up observation of EM counterparts.
The RTGW model is designed to process LISA data in this paper, but it can be easily adapted for other space-based GW detectors, such as Taiji and TianQin.

The paper is organized as follows. In Sec.~\ref{sec:model}, we describe the structure of the RTGW model. Following that, we introduce the generation of training data in Sec.~\ref{sec:dataset}. In Sec.~\ref{sec:results}, we demonstrate the reliability and effectiveness of our model across various test sets. Finally, summary and discussion are provided in Sec.~\ref{sec:summary}.

\section{Model Architecture}
\label{sec:model}
\begin{figure}[htb]
    \centering
    \includegraphics[width=1\linewidth]{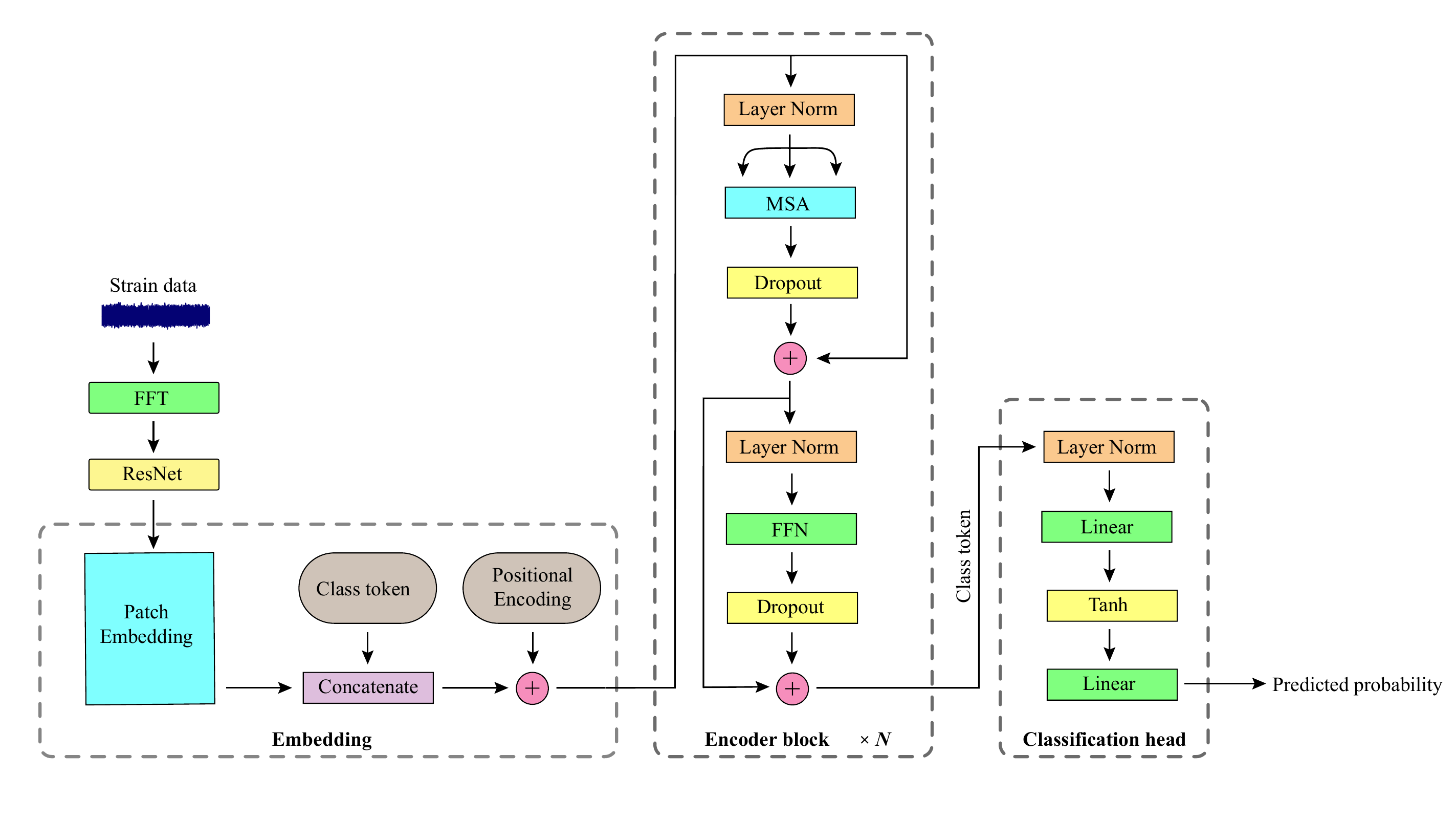}
    \caption{Framework of the RTGW model. The three modules of the Transformer architecture developed in this work, namely embedding, encoder, and classification head, are enclosed within the dashed gray box.}
    \label{fig:RTGW}
\end{figure}

For a deep learning model, the task of detecting GW signals in noisy data can be framed as a binary classification problem. 
Specifically, when presented with a segment of strain data $s$, the model categorizes it into one of two classes, where these classes represent the presence or absence of a GW signal in the given segment. 
To achieve this objective, we develop a model architecture, as illustrated in Fig.~\ref{fig:RTGW}, which seamlessly integrates ResNet and Transformer.
The architecture is inspired by the Vision Transformer~\cite{dosovitskiy:2020image}, a model designed particularly for image recognition.
In this work, the RTGW model fully leverages the capability of residual networks to extract hidden features and the capability of Transformer to deal with long-range dependencies in data.

As shown in Fig.~\ref{fig:RTGW}, the RTGW model takes time-domain data extracted from detector strain data as its initial input. 
The data segment undergoes a Fast Fourier Transform (FFT)~\cite{Cooley:1965zz} to transition into the frequency domain, and it is whitened by the power spectral density (PSD) of instrumental noise. 
This process can be regarded as data preprocessing.
Subsequently, the preprocessed data is fed into a ResNet.
The ResNet, proposed in Ref.~\cite{He_2016_CVPR}, exhibits strong feature extraction capabilities.
The residual connection structure used in it can effectively address the issue of degradation in deep networks. 
In this paper, we employ ResNet-50~\cite{He_2016_CVPR}, originally designed for image recognition, with modifications to enable it to handle one-dimensional GW data.
The ResNet ultimately outputs a feature map $\mathbf{x}_{\text{map}} \in \mathbb{R}^{C \times L}$ for analysis by subsequent layers in the network.
We fix the channel of the feature map at $C = 1024$, and the length $L$ depends on the size of the initial input. 
 
The remaining part of the RTGW model is built upon the Transformer architecture. 
The Transformer, introduced in Ref.~\cite{NIPS2017_3f5ee243} as a generative model, demonstrated its application in machine translation tasks. 
The Transformer is adept at handling long sequence data and effectively captures global features, making it successful in natural language processing.  
Additionally, it has found application in computer vision~\cite{dosovitskiy:2020image}.
In the context of space-based GW data analysis, GW signals can be observed by detectors for extended durations.
The Transformer is well-suited for capturing long-range dependencies of GW signals obscured by noise.
In this work, we construct the Transformer architecture with three modules: embedding, encoder, and classification head, as indicated by dashed gray box in Fig.~\ref{fig:RTGW}.

The embedding module takes the feature map $\mathbf{x}_{\text{map}} \in \mathbb{R}^{C \times L}$ as input. 
Firstly, in the patch-embedding layer, the feature map is reshaped into a sequence of one-dimensional patches $\mathbf{x}_p \in \mathbb{R}^{n \times (C \cdot P)}$, which can be written as
\begin{equation}
\mathbf{x}_p = [\mathbf{x}^1_p; \mathbf{x}^2_p;\cdots; \mathbf{x}^n_p] \quad  \text{for} \quad \mathbf{x}^i_p \in \mathbb{R}^{C \cdot P},
\end{equation}
where the size of patches is denoted as $P$ and the number of patches is denoted as $n = \frac{L}{P}$.
Then the patches will be mapped to the latent vector size of the encoder module (denoted as $d_{\text{enc}}$) through a trainable linear projection $\mathbf{E} \in \mathbb{R}^{ (C \cdot P) \times d_{\text{enc}}}$.
Following that, we concatenate a learnable class token, denoted as $\mathbf{x}_{\text{class}}\in \mathbb{R}^{1 \times d_{\text{enc}}}$, to the resulting patches, which is used to learn class information during the training process. 
This token's state at the output of the encoder module will be utilized for classification predictions.
Moreover, the encoder does not have an inherent sense of sequential order because it processes the patches in parallel.
In order for the Transformer to make use of the order of a sequence, it is common practice to encode relative or absolute positional information of patches within the sequence, and subsequently input this encoded information into the encoder.
In this work, we directly add a learnable positional encoding $\mathbf{E}_{\text{pos}} \in \mathbb{R}^{ (n+1) \times d_{\text{enc}}}$ to the input sequence.
Therefore, the embedding module finally outputs a sequence $\mathbf{z}_0$ that is given by
\begin{equation}
\mathbf{z}_0 = [\mathbf{x}_{\text{class}}; \mathbf{x}^1_p \mathbf{E}; \mathbf{x}^2_p \mathbf{E};\cdots; \mathbf{x}^n_p \mathbf{E}] + \mathbf{E}_{\text{pos}}.
\end{equation}

The next module in the Transformer architecture is the encoder.
As shown in Fig.~\ref{fig:RTGW}, the sequence $\mathbf{z}_0$ will pass through a stack of $N$ encoder blocks.
We employ the architecture of a standard Transformer encoder proposed in Ref.~\cite{NIPS2017_3f5ee243}, but do layer norm prior to the multi-head self-attention (MSA) mechanism and the position-wise feed-forward network (FFN).
Thus, the encoder block is composed of  
\begin{eqnarray}
\nonumber
\mathbf{z}_{j}' &=& \text{Dropout}(\text{MSA}(\text{LayerNorm}(\mathbf{z}_{j-1}))) + \mathbf{z}_{j-1}, \\
\mathbf{z}_{j} &=& \text{Dropout}(\text{FFN}(\text{LayerNorm}(\mathbf{z}_{j}'))) + \mathbf{z}_{j}',
\end{eqnarray} 
with $j = 1, 2, \cdots, N$.
Note that, the MSA is a key innovation that contributes to the Transformer's success in handling sequential data.
It allows the model to maintain sensitivity to the global context of elements when processing long sequences.
In the MSA mechanism, each element of an input sequence $\mathbf{z}$ undergoes $n_{\text{head}}$ times linear projection into $d_k$, $d_k$ and $d_v$ dimensions:
\begin{equation}
[\mathbf{Q}_i; \mathbf{K}_i;  \mathbf{V}_i] =  [\mathbf{z} \mathbf{W}_i^Q;  \mathbf{z} \mathbf{W}_i^K;  \mathbf{z} \mathbf{W}_i^V] \quad \text{for} \quad  i = 1, 2, \cdots,n_{\text{head}},
\end{equation}
where $\mathbf{W}_i^Q \in \mathbb{R}^{d_{\text{enc}} \times d_k}$, $\mathbf{W}_i^K \in \mathbb{R}^{d_{\text{enc}} \times d_k}$ and $\mathbf{W}_i^V \in \mathbb{R}^{d_{\text{enc}} \times d_v}$ are learnable parameter matrices.
Then, an output of scaled dot-product attention~\cite{NIPS2017_3f5ee243}, denoted as $\text{head}_i \in \mathbb{R}^{(n+1) \times d_v}$, is computed from
\begin{equation}
\text{head}_i = \text{Attention}(\mathbf{Q}_i; \mathbf{K}_i;  \mathbf{V}_i) = \text{Softmax} \left(\frac{\mathbf{Q}_i \mathbf{K}_i^T}{\sqrt{d_k}} \right)\mathbf{V}_i.
\end{equation}
These heads are concatenated and once again projected using a learnable parameter matrix $\mathbf{W}^O \in  \mathbb{R}^{(n_{\text{head}}\cdot d_v)\times d_{\text{enc}}}$, resulting in the final output of MSA:
\begin{equation}
\text{MSA}(\mathbf{z}) = \text{Concat}(\text{head}_1, \text{head}_2, \cdots, \text{head}_{n_{\text{head}}})\mathbf{W}^O.
\end{equation}
Furthermore, the FFN in the encoder block consists of two linear transformations with a GeLU activation~\cite{Hendrycks:2016qxa} in between, which can be presented by
\begin{equation}
\text{FFN}(\mathbf{z}) = \text{GeLU}(\mathbf{z}\mathbf{W}_1 + \mathbf{b}_1)\mathbf{W}_2 + \mathbf{b}_2,
\end{equation} 
where $\mathbf{W}_1 \in \mathbb{R}^{d_{\text{enc}} \times (4\cdot d_{\text{enc}})}$, $\mathbf{W}_2 \in \mathbb{R}^{(4\cdot d_{\text{enc}})\times d_{\text{enc}}}$, $\mathbf{b}_1 \in \mathbb{R}^{(n+1) \times (4\cdot d_{\text{enc}})}$ and $\mathbf{b}_2 \in \mathbb{R}^{(n+1) \times d_{\text{enc}}}$ are learnable parameter matrices.
Note that, the Transformer was initially designed for natural language processing as a generative model with both encoder and decoder.
However, in this work, we only need the encoder to address the classification task.

The last module of our model is the classification head which is implemented by a multi-layer perceptron (MLP)~\cite{Rumelhart:1986gxv}.
The MLP transforms the class token $\mathbf{z}^0_{N}$ extracted from the output of the last encoder block into a class prediction. 
In the binary classification task, the prediction can be expressed as two predicted probabilities through a softmax function~\cite{bishop2006pattern}, which are between 0 and 1 and their sum equals 1.
The entire RTGW model is implemented using the PyTorch framework~\cite{NEURIPS2019_9015}.

To train a classification model, the cross-entropy loss is commonly used.
It measures the difference between the predicted probabilities output by the model and the true class labels, encouraging the model to assign higher probabilities to the correct classes. 
For the binary classification task, the cross-entropy loss is given by
\begin{equation}
\mathcal{L} = -\frac{1}{N_s} \sum_{i=1}^{N_{\text{s}}} y_i\log(\hat{y}_i) + (1 - y_i)\log(1-\hat{y}_i),
\end{equation}
where $y_i$ denotes the true label of the $i$-th training sample,  $\hat{y}_i$ denotes one of the predicted probabilities and $N_s$ is the total number of samples.
We minimize the loss function using the Adam optimizer~\cite{kingma2014:adam}, enabling the RTGW model to adjust its parameters more effectively to make the predicted results closer to reality.

\section{Datasets}
\label{sec:dataset}
In this work, we consider the strain data recorded by a GW detector consists of a random noise, and possibly a GW signal.
Thus, the true label of a data sample $s(t)$ to be analyzed by the RTGW model can be set as
\begin{eqnarray}
\nonumber
\text{Label} \quad 0: \quad s(t)&=& n(t), \\
\text{Label} \quad 1: \quad s(t) &=& n(t) + h(t;\Theta),
\end{eqnarray}
where $n(t)$ denotes the random noise and $h(t;\Theta)$ denotes the GW signal with a set of physical parameters $\Theta$.
We refer to samples labeled as 0 as negative and samples labeled as 1 as positive.

Specifically, we generate GW signals using IMRPhenomD model~\cite{Husa:2015iqa,Khan:2015jqa}, which models nonprecessing spinning inspiral–merger–ringdown waveforms.
Furthermore, the time-delay interferometry (TDI) techniques~\cite{Tinto:2002de} are necessary to suppress laser frequency noise in space-based GW detectors.
In this paper, we apply the response of TDI A and E channels~\cite{Prince:2002hp} to the IMRPhenomD waveforms.
A simulated signal $h(t;\Theta)$ is characterized by 11-dimensional set of physical parameters $\{ M, q, s_{1z}, s_{2z}, d_L, t_c, \phi_c, \iota, \psi, \beta, \lambda\}$.
Here, $M$, $q$ and $(s_{1z}, s_{2z})$ denote the total mass, mass ratio and dimensionless spins of MBHB respectively, $d_L$ is the luminosity distance to the binary, $t_c$  is the coalescence time, $\phi_c$ is the coalescence phase, $\iota$ is the angle between the orbital angular momentum of the binary and the line-of-sight and $\psi$ is the polarization angle. 
The sky location of the binary is determined by the ecliptic latitude and ecliptic longitude $(\beta, \lambda)$.
In the training stage, we simulate GW signals with a fixed luminosity distance and the other 10 physical parameters randomly sampled from the prior in Tab.~\ref{tab:prior}.
In order to make the RTGW model sensitive to the inspiral phase of MBHB signal,  each training sample is generated by extracting a segment before final merger from a simulated signal.
The time interval between the segment and the final merger is uniformly randomly sampled within $[0.2 \text{d}, 2.0\text{d}]$, which ensures diversity in the training data.
Moreover, the segment is rescaled to achieve a SNR that is randomly sampled.
The prior of SNR is also given in the Tab.~\ref{tab:prior}.
We set the number of points of training data sample as 81,920 and the sampling time interval as 5 seconds.
Hence, each sample represents a 4.7-day segment of LISA data.
The MBHB signals are generated using codes provided by LISA Data Challenge group~\cite{ldcweb}.

In this work, the random noise present in strain data is composed of the instrumental noise and the confusion noise due to galactic binaries.   
The instrumental noise is simulated as a colored Gaussian noise with the design PSD of TDI A and E channels, which is given by~\cite{ldcmanual001,ldcmanual002}
\begin{equation}
S_{\text {ins}}(f) = 8 \sin ^2 \omega l\left[4\left(1+\cos \omega l+\cos ^2 \omega l\right) S_{\text{acc}}+(2+\cos \omega l) S_{\text{oms}}\right],
\end{equation}
with
\begin{equation}
\begin{aligned}
\sqrt{S_{\text {acc }}(f)} & =\frac{3 \times 10^{-15}}{2 \pi f c} \sqrt{1+\left(\frac{0.4 \times 10^{-3}}{f}\right)^2} \sqrt{1+\left(\frac{f}{8 \times 10^{-3}}\right)^4} \ \left[\frac{1}{\sqrt{\mathrm{Hz}}}\right], \\
\sqrt{S_{\mathrm{oms}}(f)} & =15 \times 10^{-12} \frac{2 \pi f}{c} \sqrt{1+\left(\frac{2 \times 10^{-3}}{f}\right)^4} \ \left[\frac{1}{\sqrt{\mathrm{Hz}}}\right].
\end{aligned}
\end{equation}
Here, $l$ represents the arm length of LISA detector, $c$ denotes the speed of light and $\omega$ is calculated as $2\pi f / c$.
Additionally, the confusion noise resulted from tens of millions of galactic binaries can be estimated by~\cite{Karnesis:2021tsh}
\begin{equation}\label{eq:conf}
S_{\text {con}} = \frac{A}{2} f^{-7 / 3} e^{-\left(f / f_1\right)^\alpha}\left(1+\tanh \left(\left(f_{\text {knee }}-f\right) / f_2\right)\right),
\end{equation}
with
\begin{eqnarray}
\nonumber
\log _{10}\left(f_1\right) &=& a_1 \log _{10}\left(T_{\text {obs }} / \text{yr}\right)+b_1, \\
\log _{10}\left(f_{\text {knee }}\right) &=& a_k \log _{10}\left(T_{\text {obs }} / \text{yr}\right)+b_k.
\end{eqnarray}   
Here, $T_{\text {obs }}$ denotes the observation duration of LISA detector and $\{A, \alpha, a_1, a_k, b_1, b_k, f_2 \}$ is a set of calibration parameters.
During the training process, we choose a fixed observation duration at $T_{\text {obs }} = 1.0 \, \text{yr}$.
The values of the calibration parameters are set as $A = 1.15 \times 10^{-44}$, $\alpha = 1.56$, $a_1 = -0.15$, $a_k = -0.37$, $b_1 = -2.72$, $b_k = -2.49$, and $f_2 = 0.00067$, which are given in Ref.~\cite{Karnesis:2021tsh}.
The confusion noise is simulated from the Eq.~\eqref{eq:conf}, with modulations resulted from TDI channels.
The total noise $n(t)$ in a training sample is the sum of the instrumental noise and the confusion noise.
We generate the two components of noise using the PyCBC package~\cite{alex_nitz_2021_5347736}.

\begin{table*}[h]
    \centering
    \renewcommand{\arraystretch}{1.2}
    \setlength{\tabcolsep}{7mm}{
    \begin{tabular}{cc}
        \hline
        Parameter & Prior \\ \hline
        $M$   & $\text{LogUniform}[10^5 M_{\odot}, 10^7 M_{\odot}]$   \\
        $q$     & $\text{Uniform}[1, 10]$   \\
        $t_c$   & $\text{Uniform}[3 \text{d}, 365 \text{d}]$   \\
        SNR   & $\text{Uniform}[15, 60]$   \\
        $(s_{1z}, s_{2z})$ & $\text{Uniform}[-1, 1]$   \\
        $\cos \iota$ & $\text{Uniform}[-1, 1]$   \\
        $\sin \beta$ & $\text{Uniform}[-1, 1]$   \\
        $\lambda$ & $\text{Uniform}[0, 2\pi]$   \\
        $\phi_c$ & $\text{Uniform}[0, 2\pi]$   \\
        $\psi$ & $\text{Uniform}[0, \pi]$      \\ \hline
    \end{tabular}}
    \caption{Prior parameter distribution for simulated GW signals.}
    \label{tab:prior}
\end{table*}

\section{Results}
\label{sec:results}
\begin{figure}[htb]
\centering
\includegraphics[width=8cm,height=6cm]{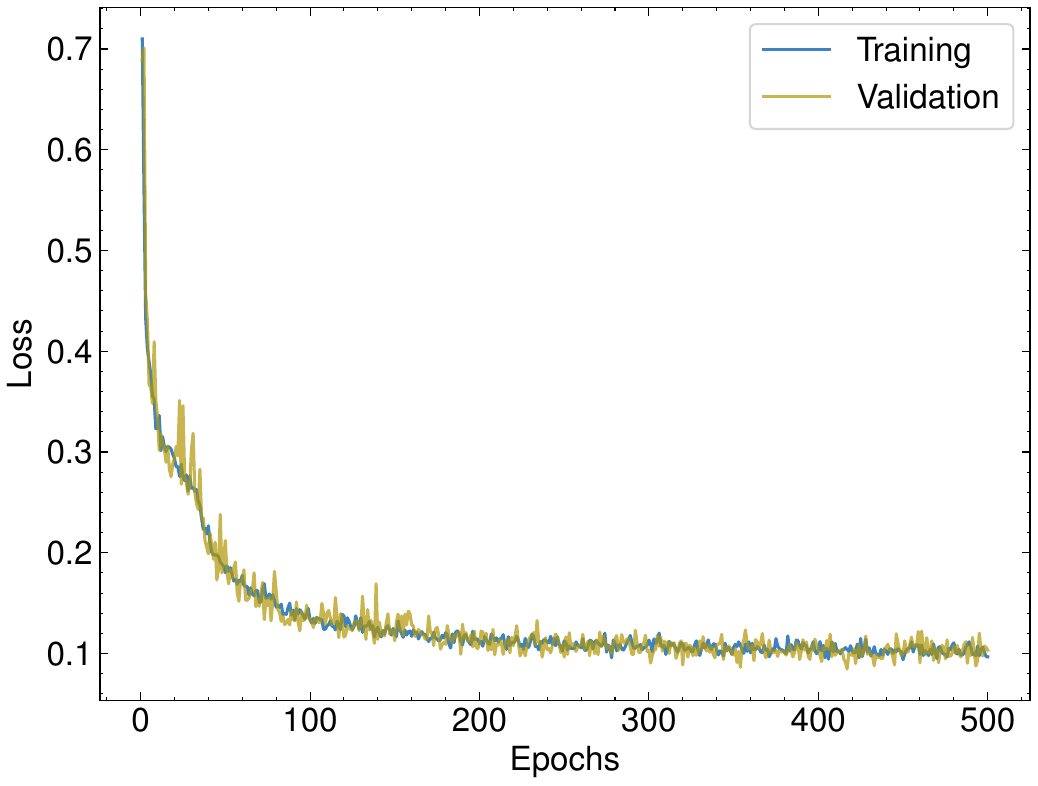}
\caption{Evolution of the losses for the training and validation sets. The training process is stopped at epoch 500.}
\label{fig:loss}
\end{figure}

In this work, all the training and testing process are conducted on an NVIDIA Tesla A40 GPU.
We generate 10,000 samples as the training set and 3,000 samples as the validation set in each epoch of the training process, which is effective in preventing overfitting.
Both the training and validation sets have half of the samples containing only random noise (negative samples), while the other half contain noise and MBHB signal (positive samples).
We train models with different selections of the hyperparameters, and evaluate their performance based on the final validation loss.
The hyperparameters are ultimately determined as the patch size $P = 2$, the number of encoder blocks $N = 6$, the latent vector size of encoder block $d_{\text{enc}} = 768$, and the number of heads $n_{\text{head}} = 12$.
Furthermore, the learning rate gradually decreases due to cosine annealing~\cite{Loshchilov:2016tgk} with a beginning at 0.00008.
The training process of the RTGW model, which takes nearly three days, is stopped at epoch 500. The losses for the training and validation sets are shown in Fig.~\ref{fig:loss}.
The trained model can analyze a segment of simulated LISA data spanning 4.7 days within 0.01 seconds.

\subsection{Tests on independent data segments}

\begin{figure}[!htbp]
\begin{center}
\subfigure[]{
\includegraphics[width=8cm,height=6cm]{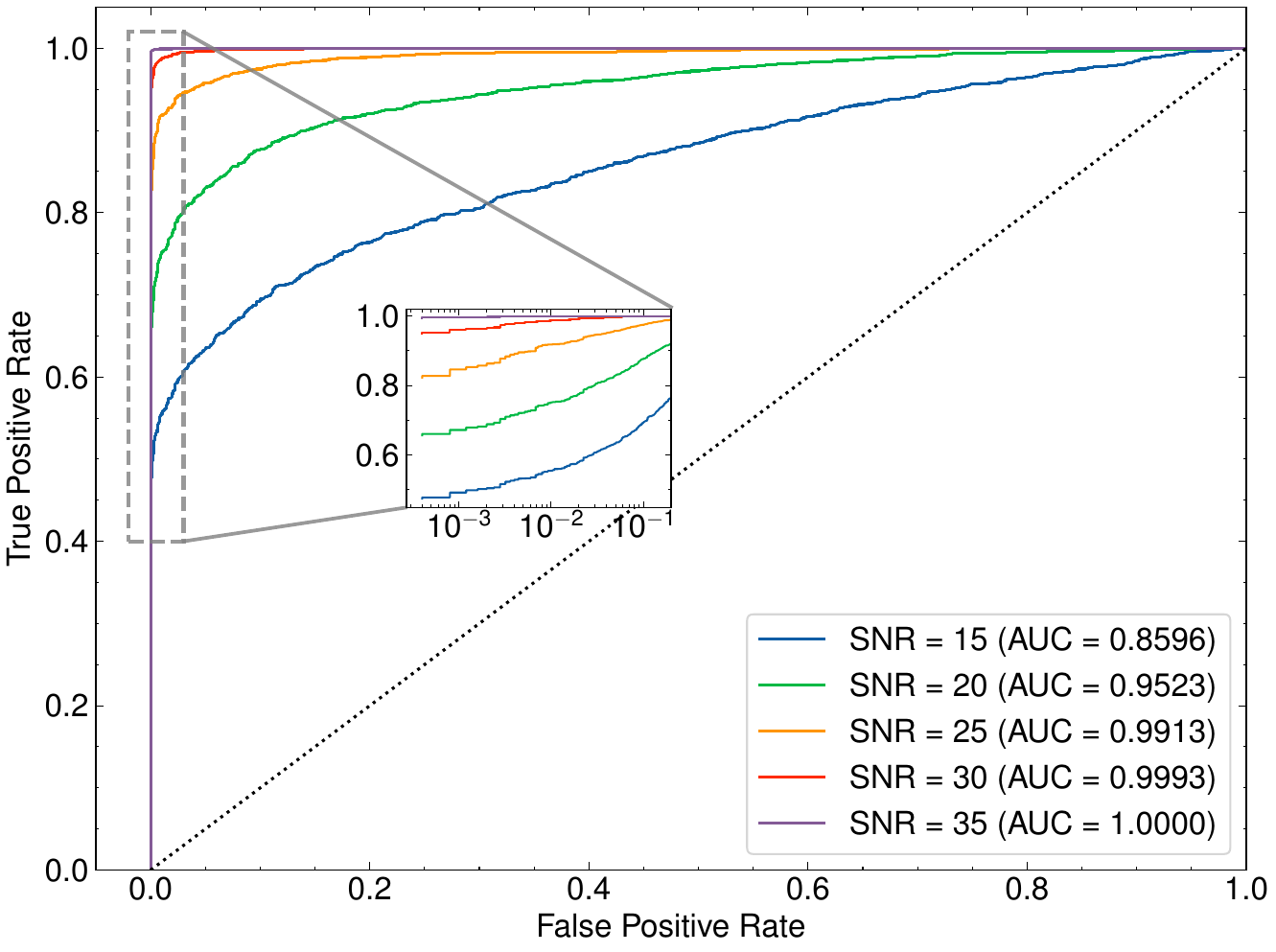}
\label{fig:roc_all}
}
\subfigure[]{
\includegraphics[width=8cm,height=6cm]{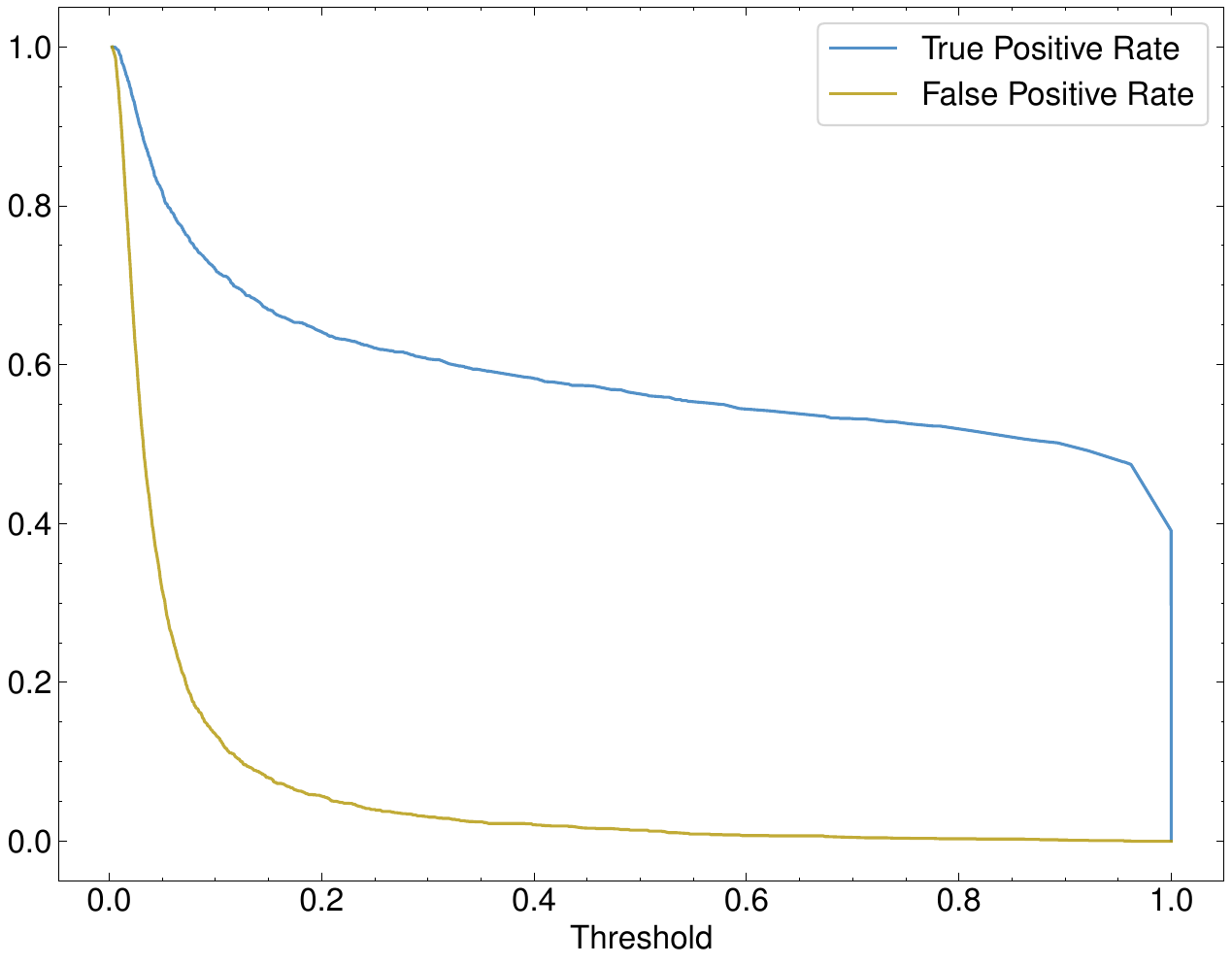}
\label{fig:threshold}
}
\caption{ROC analysis for the RTGW model. (a) The ROC curves for test datasets with different SNRs. 
Each test dataset comprises 5,000 samples, with an equal split between positive and negative samples. 
The positive samples contain injected segments of MBHB signals simulated with fixed SNRs of 15, 20, 25, 30 and 35.
The black dotted line denotes the line of random classifier.
(b) The FPR and TPR as functions of the threshold.
The TPR is obtained from the test dataset with a fixed SNR of 15.
}
\label{fig:roc}
\end{center}
\end{figure}

To evaluate the sensitivity of our model to MBHB signals, we test it on datasets with different SNRs.
Specifically, each dataset consists of 5,000 test samples, with an equal split between positive and negative samples.
The injected signals in positive samples are simulated at a fixed SNR of 15, 20, 25, 30, and 35, respectively.
The other physical parameters are randomly sampled from the prior in Tab.~\ref{tab:prior}, which is the same with the training set.
The performances of our model on these datasets are visualized by the receiver operating characteristic (ROC) curves~\cite{FAWCETT2006861} shown in Fig.~\ref{fig:roc_all}.
The ROC curve depicts the true positive rate (TPR) against the false positive rate (FPR) for various threshold values applied to the predicted probabilities produced by the model.
Moreover, the area under the ROC curve (AUC)~\cite{FAWCETT2006861,BRADLEY19971145,hanley1982meaning} can be used as a numerical summary of the model's performance, with AUC closer to 1 indicating better performance.
As illustrated in Fig.~\ref{fig:roc_all}, the model performs better on the datasets with a higher SNR.
With the FPR fixed at 0.01, our model achieves a TPR of over 0.9 on datasets with a SNR above 25.
As the SNR decreases to 15, the TPR drops to 0.55.
The RTGW model exhibits high sensitivity to MBHB signals from the premerger stages while keeping a low false alarm, indicating its potential to detect signals when the SNR accumulates to 15.
In practice, a predefined threshold is required to determine the decision boundary between the presence and absence of MBHB signals.
Fig.~\ref{fig:threshold} shows the FPR and the TPR as functions of the threshold.
The TPR is obtained from the test dataset with a fixed SNR of 15.
The results indicate that the false alarm rate decreases as the threshold increases, but the model becomes less sensitive to MBHB signals. 

\begin{figure}[htb]
\centering
\includegraphics[width=8cm,height=6cm]{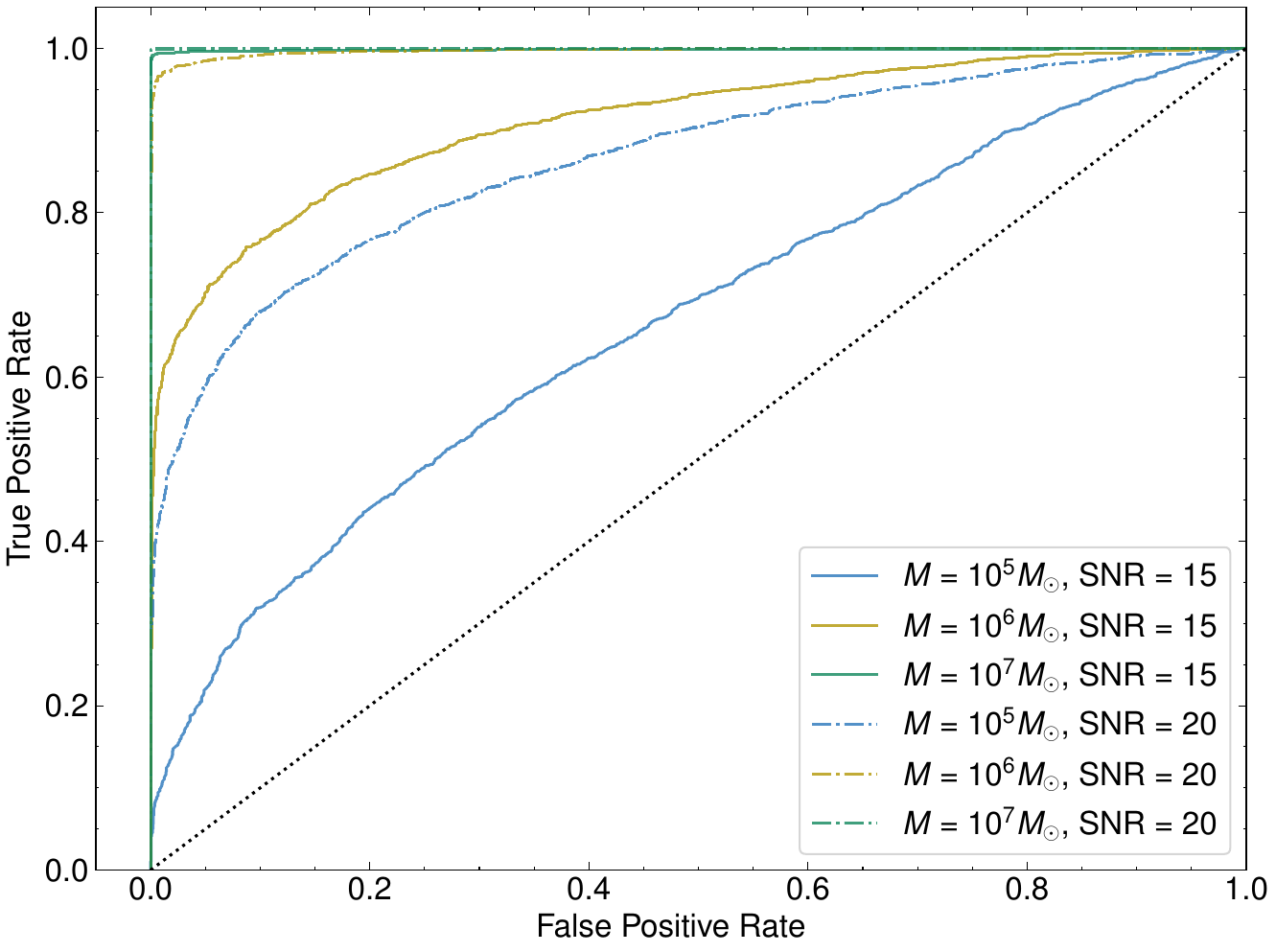}
\caption{The ROC curves for test datasets with different total masses and SNRs. Each test dataset comprises 5,000 samples, with an equal split between positive and negative samples. The positive samples contain injected segments of MBHB signals simulated with fixed total masses and SNRs of $(10^5 M_{\odot}, 15)$, $(10^6 M_{\odot}, 15)$, $(10^7 M_{\odot}, 15)$, $(10^5 M_{\odot}, 20)$, $(10^6 M_{\odot}, 20)$, and $(10^7 M_{\odot}, 20)$. The black dotted line denotes the line of random classifier.}
\label{fig:roc_mass}
\end{figure}

Furthermore, the RTGW model demonstrates varying sensitivity to MBHB signals with different total masses.
We regenerate the datasets in the same manner, with the only difference being that the total mass of the MBHBs is fixed at $10^5 M_{\odot}$, $10^6 M_{\odot}$, and $10^7 M_{\odot}$, respectively.
The ROC curves obtained from tests on datasets with SNR of 15 and 20 are depicted in Fig.~\ref{fig:roc_mass}.
The results indicate that, when SNR is fixed, our model is more sensitive to signals with larger total masses, which typically reside in lower frequency bands.

\subsection{Tests on continuous signals}

\begin{figure}[htb]
    \centering
    \includegraphics[width=16cm,height=9cm]{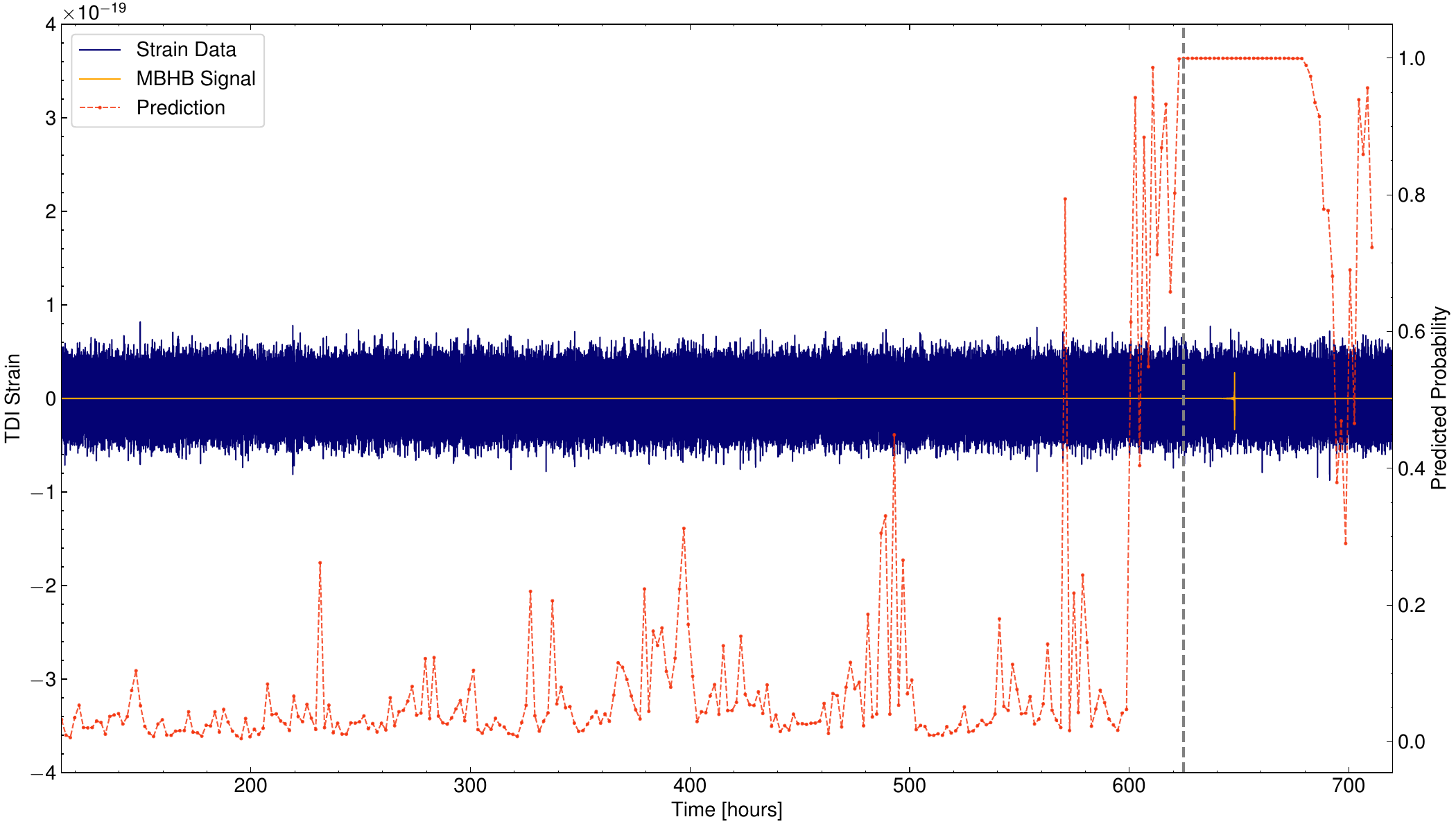}
    \caption{Performance of the RTGW model on a 30-day simulated strain data. The noisy data of TDI A channel is denoted by the blue line and the injected MBHB signal is denoted by the orange line. The signal covers the inspiral, merger and ringdown phases, with an SNR set to 200. The red points represent the model's predicted probabilities for label 1, with a time interval of 2 hours between adjacent points. The model detects the signal at the time marked by the gray dashed line utilizing a threshold of 0.999.}
    \label{fig:single}
\end{figure}

In both training and the aforementioned testing, all samples provided to the RTGW model are independent data segments.
However, in real data analysis, we need to deal with continuously lengthening sequences as the accumulation of observation time.
As mentioned in Sec.~\ref{sec:dataset}, our model takes input data segment which covers an observational time of about 4.7 days for a detector.
The model is capable of processing the data segment within 0.01 seconds.
Therefore, as the detector's strain data updates, our model enables near real-time analysis. 
It sequentially processes segments extracted from the updated data stream, producing a series of predicted probabilities.
The identification of a MBHB signal can be triggered by setting a threshold on the predicted probability.
Specifically, in this work, if the predicted probability for label 1 exceeds 0.999, we classify it as indicating the presence of a MBHB signal in the strain data.
The high threshold ensures a low false alarm rate but compromises the model's sensitivity to MBHB signals.
Fig.~\ref{fig:single} shows the model's predictions on a 30-day simulated LISA data, which contains a GW signal from a MBHB with a total mass of $10^6 M_{\odot}$ and a mass ratio of 5.
The signal covers the inspiral, merger, and ringdown phases, and its SNR is set to 200. 
The time interval between two consecutive predictions of the RTGW model is set to 2 hours.
Based on a threshold of 0.999, our model can identify the MBHB signal approximately 23.3 hours before the final merger, with an accumulated SNR reaching 21.0.

\begin{figure}[!htbp]
\begin{center}
\subfigure[]{
\includegraphics[width=8cm,height=6cm]{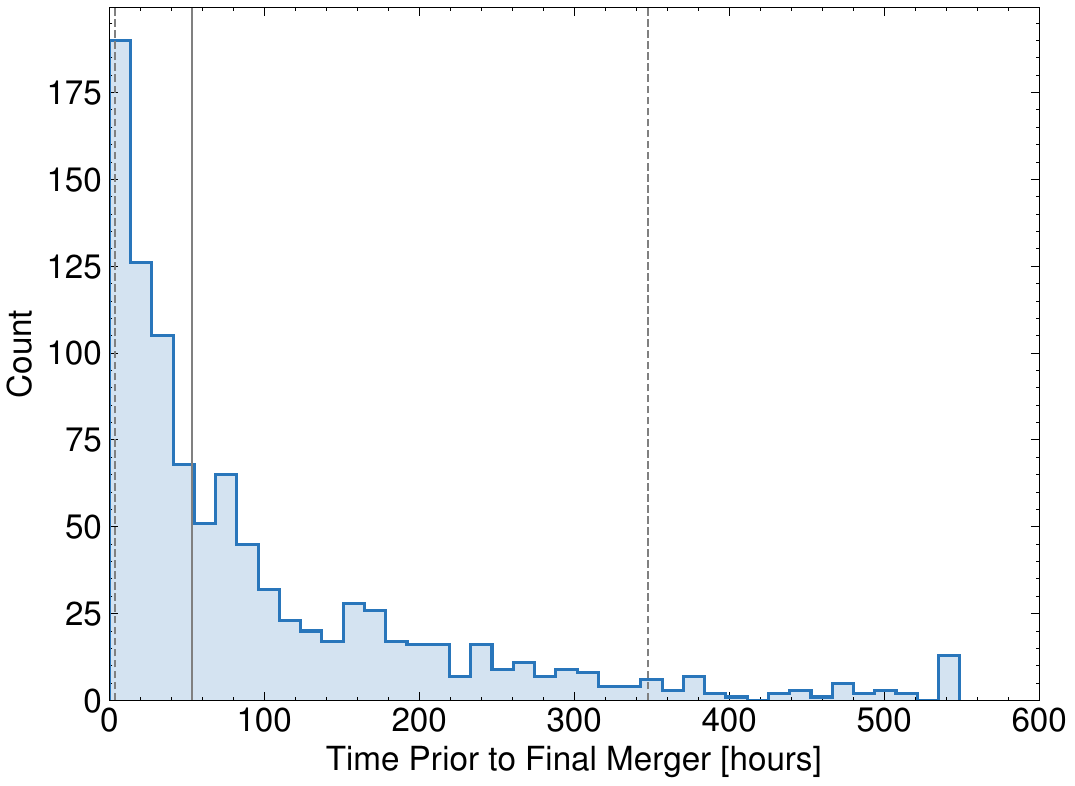}
\label{fig:time_prior}
}
\subfigure[]{
\includegraphics[width=8cm,height=6cm]{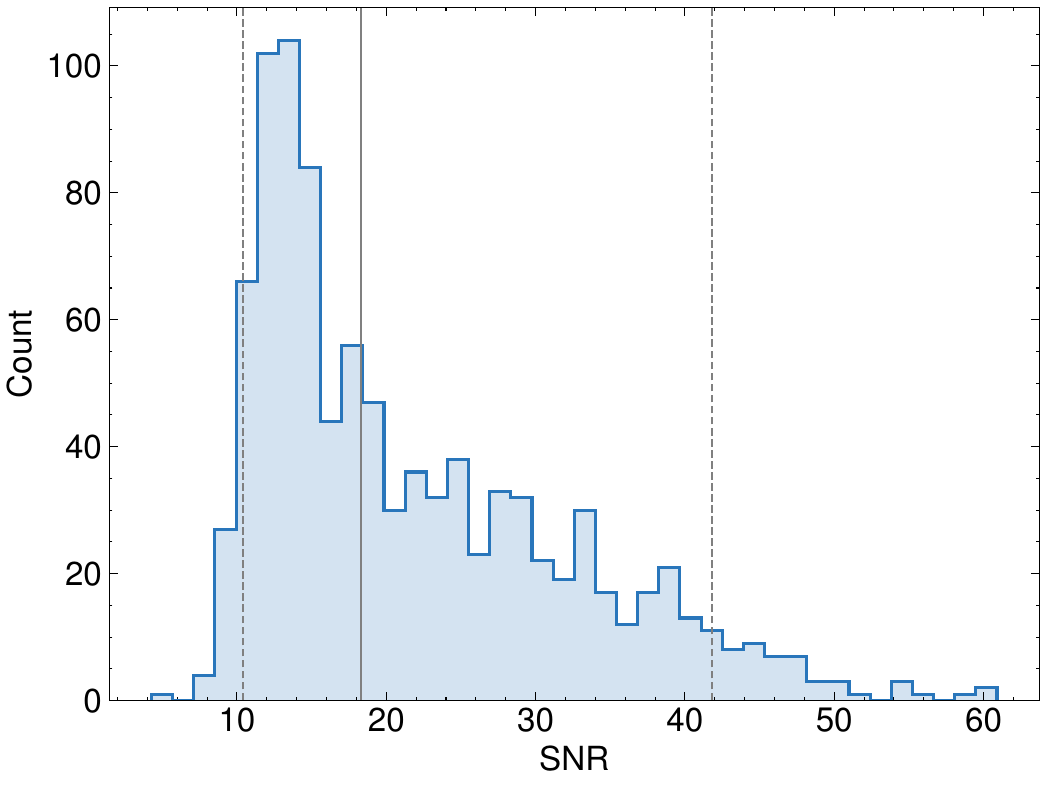}
\label{fig:snr_all}
}
\caption{Detect capability of the RTGW model. The model is tested on a dataset comprising 1,000 positive samples and 1,000 negative samples. Each sample represents a 30-day simulated strain data. All the injected MBHB signals in positive samples cover the inspiral, merger and ringdown phases and their SNRs are uniformly sampled from the range of $[100, 500]$. The subplot (a) shows the time remaining until the final merger when these signals are identified by the model, and their accumulated SNRs at the time of identification are shown in subplot (b). The area between the two gray dashed lines represents the results of $90\%$ of the samples, while the gray solid line denotes the median.}
\label{fig:local_time}
\end{center}
\end{figure}

More generally, we generate a test set consisting of 2,000 samples to evaluate the model's performance on long-duration data.
Half of these samples are noise-only, while the other half are a mixture of noise and MBHB signal.
Each sample represents a 30-day simulated strain data.
Similar to the MBHB signal shown in Fig.~\ref{fig:single},  the signals generated for this test set also cover the inspiral, merger, and ringdown phases, with their SNRs uniformly randomly sampled from the range of $[100, 500]$.
The values of other physical parameters are sampled from the prior given in Tab.~\ref{tab:prior}.
In the test, we also choose a threshold of 0.999 for the model's predicted probabilities.
Our model can identify $97.4\%$ of the MBHB signals before the final merger, and achieve an FPR of $0.5\%$.
This indicates that the threshold enables the model to maintain a low false alarm while achieving a high sensitivity to MBHB signals.
Furthermore, Fig.~\ref{fig:time_prior} illustrates the time prior to the final merger at which the MBHB signals are identified by our model.
Considering $90\%$ of the samples, the signals are detected $53.5_{-49.8}^{+293.9}$ hours before the final merger.
At the moment of being detected, the SNRs of these signals reach $18.3_{-7.9}^{+23.5}$, as depicted in Fig.~\ref{fig:snr_all}.
Hence, the RTGW model demonstrates the ability to detect MBHB signals with low SNR before the final merger, thereby providing essential lead time for subsequent EM observations.

\begin{figure}[htb]
\centering
\includegraphics[width=8cm,height=6cm]{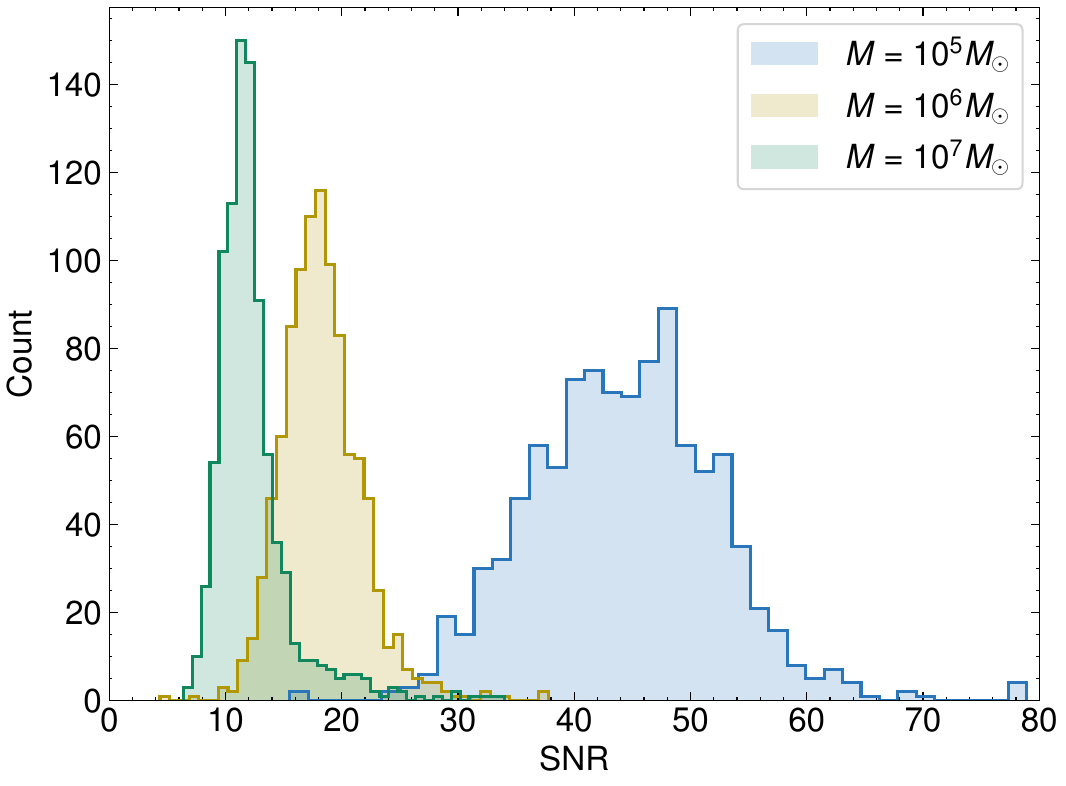}
\caption{Distribution of SNRs at the time of identification for MBHB signals with different total masses. The RTGW model is tested on three datasets generated with fixed total masses of $10^5 M_{\odot}$, $10^6 M_{\odot}$, and $10^7 M_{\odot}$, respectively.}
\label{fig:snr_mass}
\end{figure}

The results shown in Fig.~\ref{fig:roc} indicate that the sensitivity of our model depends on the total mass of MBHB.
To assess the model's detection boundary concerning various total masses, we generated datasets with total masses fixed at $10^5 M_{\odot}$, $10^6 M_{\odot}$, and $10^7 M_{\odot}$, each containing 1,000 samples.
The samples here are the same as those employed in the test shown in Fig.~\ref{fig:local_time}, consisting of a mixture of noise and MBHB signal spanning a 30-day duration, with the additional physical parameters sampled from the same prior.
As shown in Fig.~\ref{fig:snr_mass}, for $90\%$ of the samples, the model detects the signals when the SNRs accumulate to $44.3_{-12.8}^{+12.3}$ for MBHBs with a total mass of $10^5 M_{\odot}$.
Correspondingly, for MBHBs with total masses of $10^6 M_{\odot}$ and $10^7 M_{\odot}$, the required SNRs are $18.0_{-4.7}^{+6.0}$ and $11.7_{-2.9}^{+6.8}$, respectively.
It suggests that the RTGW model is less sensitive to signals from MBHBs with smaller total masses.
However, this does not imply that our model will only detect them when the final merger approaches more closely.
Actually, due to variations in noise PSD at different frequencies, GW signals simulated from a total mass of  $10^5 M_{\odot}$ accumulate a significant SNR earlier in the inspiral phase. 
Thus, despite the model having lower sensitivity to these signals, it can still detect them well in advance of the final merger.
In $90\%$ of the test cases, our model identifies the signals $255.3_{-195.6}^{+293.2}$ hours before the final merger.
For MBHBs with total masses of $10^6 M_{\odot}$ and $10^7 M_{\odot}$, the lead times are $59.6_{-50.0}^{+129.9}$ hours and $9.7_{-8.4}^{+25.8}$ hours, respectively.

\section{Summary and Discussion}
\label{sec:summary}
In this study, we have developed an effective deep learning approach for identifying MBHB signals for space-based GW observations.
Compared to the convolutional neural network based model in Ref.~\cite{Ruan:2021fxq}, we employ the Transformer architecture in the RTGW model, which contributes to effectively capturing the global dependencies of GW signals.
The MSA mechanism in the Transformer architecture dynamically adjusts attention weights based on different parts of input sequence, enabling the RTGW model to better focus on the relevant parts of signal and mitigate the impact of noise.
Therefore, the model has ability to identify signals even under low SNR conditions.  

The computational cost of a deep learning model predominantly stems from the iterative optimization process, which entails training on extensive data.
The RTGW model, once trained, is capable of processing a segment of LISA data spanning 4.7 days within 0.01 seconds.
Furthermore, the model demonstrates high sensitivity to GW signals from MBHBs while maintaining a low false alarm.
Based on its predictions, it is expected to detect the signals several hours to over ten days prior to the final merger. 
This premerger detection lays the foundation for the rapid localization of MBHBs, which is crucial for the observation of EM signatures appearing before, during or immediately after the merger. 
The coincident GW and EM detection of MBHBs provides a new avenue for multi-messenger astronomy.

The confusion noise, formed by GW signals from numerous galactic binaries, is simulated as Gaussian and stationary in this work.
However, in real-world scenarios, it is anticipated to manifest non-Gaussian behavior and evolve over time.
In future work, the RTGW model is supposed to be trained with simulated noise that more closely resembles real-world conditions, enabling it to accommodate authentic noise patterns.
Besides MBHBs, some other sources are also detectable for space-based GW detectors, such as extreme mass-ratio inspirals~\cite{Amaro-Seoane_2007} and possible intermediate mass black hole binaries~\cite{PortegiesZwart:2002iks}.
GW signals from these sources may overlap with MBHB signals, which are not considered in this study.
In addition, real data may exhibit disturbances like glitches and data gaps~\cite{Cornish:2021smq}.
Our model has successfully learned to capture the essential characteristics of MBHB signals.
The pre-trained model should be fine-tuned through transfer learning to leverage existing knowledge in adapting to new features present in real-world data.

\begin{acknowledgments}

This work is supported by the National Natural Science Foundation of China Grant No. 12247140, No. 12075297 and No. 12235019.
We thank the LISA Data Challenge group to provide the software.

\end{acknowledgments}

\bibliographystyle{apsrev4-2}
\bibliography{ref}

\end{document}